# Evidence of a hydrated mineral enriched in water and ammonium molecules in the Chang'e-5 lunar sample

**Authors:** Shifeng Jin[1,2]†, Munan Hao[1,2]†, Zhongnan Guo[3], Bohao Yin[4], Yuxin Ma[1], Lijun Deng[5], Xu Chen[1], Yanpeng Song[1], Cheng Cao[1,2], Congcong Chai[1,2], Qi Wei[6], Yunqi Ma[6], Jiangang Guo[1,7], Xiaolong Chen[1,2,7]*

**Affiliations:**

[1]Beijing National Laboratory for Condensed Matter Physics and Institute of Physics, Chinese Academy of Sciences; Beijing 100190, China.

[2]School of Physical Sciences, University of Chinese Academy of Sciences, Beijing 100190, China.

[3]School of Chemistry and Biological Engineering, University of Science & Technology Beijing, Beijing 100083, China.

[4]School of Earth System Science, Tianjin University, Tianjin 300072, China

[5]Center of Advanced Analysis & Gene Sequencing, Zhengzhou University, Zhengzhou 450001, China

[6]Qinghai Institute of Salt Lakes, Chinese Academy of Sciences, Xining 810008, China

[7]Songshan Lake Materials Laboratory, Dongguan 523808, China

*Corresponding author. Email: chenx29@iphy.ac.cn

† These authors contributed equally to this work.

**Abstract**

The presence and distribution of water on the Moon are fundamental to our understanding of the Earth-Moon system. Despite extensive research and remote detection, the origin and chemical form of lunar water ($H_2O$) have remained elusive. In this study, we present the discovery of a hydrated mineral, $(NH_4)MgCl_3·6H_2O$, in lunar soil samples returned by the Chang'e-5 mission, containing approximately 41 wt% $H_2O$. The mineral's structure and composition closely resemble novograblenovite, a terrestrial fumarole mineral formed through the reaction of hot basalt with water-rich volcanic gases, and carnallite, an earth evaporite mineral. We rule out terrestrial contamination or rocket exhaust as the origin of this hydrate, based on its chemical and isotopic compositions and formation conditions. The presence of ammonium indicates a more complex lunar degassing history and highlights its potential as a resource for lunar habitation. Our findings also suggest that water molecules can persist in sunlit areas of the Moon as hydrated salt, providing crucial constraints to the fugacity of water and ammonia vapor in lunar volcanic gases.

# Main

The Earth and Moon are often regarded as nonidentical twins, believed to have originated from a shared giant impact event (*1*). On the Earth, water plays a crucial role in planetary evolution through the influence on melting, viscosity, magma crystallization and volcanic eruption (*2,3*). However, initial analyses of lunar samples returned by the Apollo and Luna missions led to the prevailing notion of a dry Moon (*4*), which imposed significant constraints on our understanding of the Moon's formation through a giant impact (*5*), the existence of a lunar magma ocean (*6*), and subsequent volcanic activity (*7*). In recent years, advancements in microanalysis techniques have challenged the concept of a "dry Moon" through the detection of small amounts of hydroxyl ($OH^-$) in various lunar samples (*8*), including pyroclastic glass beads (*9*), melt inclusions (*10*), apatite (*11*), and anorthosite (12). Estimates of water abundances in the lunar mantle source regions vary widely, ranging from 0.3 to 1410 ppm (*13-14*), with the highest values comparable to those found in mid-oceanic ridge basalts (MORB) on Earth (*10*). Additionally, remote sensing and direct impact missions have confirmed the presence of water molecules ($H_2O$) in permanently shadowed craters near the lunar poles (*15,16*). The significant amount of water molecules detected, approximately 5.6% by weight, is commonly believed to be in the form of water-ice (*17*), originating from exogenous sources such as interactions with the solar wind and delivery through impacts (*8*). Due to the low oxygen fugacity $f(O_2)$ on the Moon (*18*), the release of $H_2$ vapors is proposed more favored over $H_2O$ vapors during lunar magma degassing (*19*), and it remains unclear whether volcanoes emitting gases have contributed to the lunar hydrosphere (*20*). Furthermore, the recent discovery of $H_2O$ on the sunlit regions of high latitudes suggests the existence of lunar water ($H_2O$) in forms other than water-ice, enabling its survival against thermal conditions (*21*). Due to the lack of returned lunar samples from high-latitude and polar regions, neither the origin nor the actual chemical form of lunar hydrogen ($H_2O$) have been determined. The Chang'e-5 (CE5) mission successfully returned 1.731 kg of lunar soil samples from northern Oceanus Procellarum at 43.058 °N (*22*), the landing site is much higher in latitude than the Apollo and Luna missions. The mare basalt from CE5 regolith have been dated at 2.0 Ga using the lead (Pb)–Pb isotope isochron technique. (*23,24*). Both in-situ reflectance spectra and examination of the returned lunar regolith revealed the presence of trace water ($OH^-$) (*25-27*), which is mostly attributed to solar wind implantation (*25*) or hydroxyl-containing apatite (*27*). We singled out more than 1000 mineral clasts (larger than 50 μm) from Sample CE5C0400 (1.5 g, Fig. 1A) by aid of an optical microscope for Raman measurement, crystallographic study and chemical analyses. Pyroxene, anorthite, olivine, ilmenite and glassy beads make up the majority of the sample along with a small amount of silica and spinel are ascertained (*28*). Noticeably, a prismatic, plate-like transparent crystal was identified to be an unknown lunar mineral (ULM-1) with a size of about 160×109×73 μm (Fig. 1B).

The chemical compositions of ULM-1 were analyzed on fragments of the crystal using both SEM energy dispersive spectroscopy (EDS) and Electron probe microanalysis (EPMA). As shown in Fig. 1C and Supplementary Fig. 1, the $K_\alpha$ lines of the EDS spectrum suggest the mineral contain Mg and abundant volatile elements N, Cl and O, see the measurement details and the choice of standards in Support Information. The analytical results are listed in Supplementary Table 1, revealing the ratios of major elements is close to N:Mg:Cl:O =1:1:3:6. Significantly, upon repeated measurements of the crystal, there is a rapid decrease in the concentration of N, Cl, and O elements compared to Mg. Meanwhile, numerous fractures appeared in the petrographic image (Supplementary Fig. 2), suggesting that the ULM-1 crystal is undergoing decomposition due to electron beam radiation. More detailed EDS & EPMA measurements revealed the existence of a variety of minor species with concentrations about 1% or more, including K, Cs, Na, Si, Al, Ca, S, and others, many of which are volatile elements (Supplementary Fig. 3-6). Upon examination of different regions in ULM-1, it was found that the spatial distribution of these minor elements is highly heterogeneous, even at the micrometer scale (Supplementary Fig. 3&4). Fine-scale variations in the composition may suggest either disequilibrium or rapidly changing equilibrium during its crystallization, or the influence of secondary processes.

The presence of water and ammonium molecule in ULM-1 crystal are confirmed in both the Raman and Infrared (IR) spectra. The strong normal vibrational modes at 3430 cm$^{-1}$ (symmetric stretching $v_1$) and at 1645 cm$^{-1}$ (bending O–H deformations $v_2$) of $H_2O$ molecule (*29*) are clearly observed in the Raman spectrum of ULM-1 as shown in Fig. 1D. Normal modes of the $NH^{4+}$ ion at 3080 cm$^{-1}$ and 1415 cm$^{-1}$ is due to its $v_1$ in phase stretching and $v_4$ triply degenerate bending, respectively (*30*). The band at 3250 cm$^{-1}$ corresponds to $NH^{4+}$ stretching vibrations (*30*). The infrared spectrum of ULM-1 shows clearly the major modes for water molecule and $NH^{4+}$, see Fig. 1E. Specifically, the IR absorption at 1650 cm$^{-1}$ can be attributed to the bending mode of water molecule, which was recently observed in the sunlit moon by SOFIA remote sensing (*21*). Overall, the Raman and IR spectra as well as the determined chemical compositions on ULM-1 are very consistent to the observations in novograblenovite $(NH_4,K)MgCl_3·6H_2O$ (*31,32*), an earth mineral recently found on basaltic lava from the 2012–2013 Tolbachik fissure eruption at the Plosky Tolbachik volcano (*32*).

The crystal structure of ULM-1 was determined by X-ray single crystal diffraction data (Supplementary Table 2). As shown in Fig. 2A&B, the mineral crystallizes in monoclinic space group C2/c, with lattice parameters $a$=9.3222(10) Å, $b$=9.5731(10) Å, $c$= 13.3328(12) Å and $\beta$=90.101(5) °, $Z$=4. Notably, the six independent H atoms on the water molecules can be clearly revealed in difference-Fourier maps, see Fig. 2c. Meanwhile, the H atoms in ammonium groups were also revealed and found to be disordered. Subsequent structure refinements revealed nearly full occupancy of the major elements N, Mg, Cl and O, and the trace elements determined by EPMA were complemented to the structure during the last refinements. The final structure parameters and the selected bond lengths and angles are shown in Supplementary Table 2-5, respectively. The chemical formula for the ULM-1 crystal can then be best represented as $[(NH_4)_{0.87}Na_{0.009}K_{0.021}Cs_{0.012}][Mg_{0.97}Ca_{0.023}Al_{0.007}]$ $Cl_3·6H_2O$ according to the results of chemical analysis, structure refinement and the principle of charge neutrality.

In the determined crystal structure, each $Mg^{2+}$ ion is coordinated by 6 O from water molecules to form a nearly regular octahedron $[Mg(H_2O)_6]^{2+}$; each $(NH_4)^+$ is connected by 6 $Cl^-$ anions through H-Cl hydrogen bonds to form corner shared octahedrons in a 3D framework (Fig. 2B). The crystal structure of ULM-1 is isostructural to novograblenovite (*31,32*). Additionally, it shares topological similarities with that of carnallite (*33*), where nearly all the $K^+$ cations in carnallite are replaced by $(NH_4)^+$ molecules. As shown in Fig. 2A, the main units of *A*, *B* and *O* in ULM-1 are inter-connected entirely by hydrogen bonds (short dotted line). Each chlorine atom accommodates six hydrogen bonds: four from O atoms at the vertices of $Mg(H_2O)_6$ octahedra and two from the ammonium ions. The O–H⋯Cl hydrogen bonds lie in the interval 2.33(2)–2.49(2) Å, while N–H⋯Cl bonds are normally longer from 2.33(1) to 2.88(1) Å (Supplementary Table 5 ). Our chemical and structural analyses suggest that ULM-1 be an extremely water (41.7 % wt.) and ammonia (6.6 % wt.) rich mineral. The occurrence of lunar hydrate is surprising, as previous studies have shown that lunar mineralogy is characterized by the absence of water-bearing minerals like clays, micas, and amphiboles (*18*). We'd like to consider three potential sources for this hydrate, obtained from CE5 lunar soils. Firstly, it may have an intriguing lunar origin as a rare fumarole mineral. Secondly, it could be a result of contamination from terrestrial sources. Lastly, it may be a chemical reaction product resulting from the interaction of rocket exhaust with lunar soils during landing. We argue, based on comprehensive chemical, structural and isotopic analysis, that this unusual CE5 hydrate should have an indigenous lunar origin.

As well known, rocks from the Moon and Earth have very similar isotopic ratios for the most abundant elements, but one of the most significant exceptions is lunar chlorine (*34,35*). Lunar $\delta^{37}Cl$ values [defined as part per thousand variations in $^{37}Cl/^{35}Cl$, relative to standard mean ocean chlorine] are determined to be in a range from -4 ‰ to +81 ‰ (*36,37*), far more spread and higher than observed in any other planetary bodies in solar system (*34-38*). The $\delta^{37}Cl$ values for terrestrial minerals, for instance, only cluster around 0 ‰ with a typical spread of ± 1.0 ‰ (*35*). Exceptionally high $\delta^{37}Cl$ values exceeding +7 ‰ are only observed in some volcanic gases (*35*). In the case of other extraterrestrial minerals, only samples from the Vesta meteorite group and comet coma exhibit $\delta^{37}Cl$ values higher than +9 ‰ (*39-41*), see Fig. 3. The chlorine isotopic composition of ULM-1 crystal is analyzed by CAMECA NanoSIMS 50L secondary ion mass spectrometer. To calibrate the matrix effects and instrument mass fractionation (IMF), terrestrial $(NH_4)MgCl_3 \cdot 6H_2O$ crystals is synthesized from aqueous solution, employing the chemical reagents $NH_4Cl$, $MgCl_2 \cdot 6H_2O$ purchased from Sinopharm Co., Ltd. The Cl isotopic ratios in the synthetic $(NH_4)MgCl_3 \cdot 6H_2O$ standard is determined based on high accuracy solid thermal ionization mass spectrometer (*42*), and the $\delta^{37}Cl$ value is determined to be +0.08(1) ‰. Relative to such a small $\delta^{37}Cl$ value in terrestrial sample, the measured $\delta^{37}Cl$ value of ULM-1 is exceptionally high, up to +24.5 ‰ on average, see Fig. 3 and Supplementary Table 6 for measurement details. As shown in Figure 3, the extremely high $\delta^{37}Cl$ value in ULM-1 is significantly higher than the minerals from terrestrial environments (*35*), instead, this $\delta^{37}Cl$ value well fall into the range of lunar minerals (*36,37*).

The determined structure of ULM-1 is isostructural to a terrestrial mineral novograblenovite $(K,NH_4)MgCl_3 \cdot 6H_2O$ (*31,32*) and closely resembles carnallite (*33*), which also shed light on the origin of this hydrate. On the Earth, novograblenovite is a rare fumarole mineral that was discovered in Tolbacik volcano only recently. In-depth analysis suggests this mineral should be formed by a reaction of basalt with hot volcanic gases enriched in $H_2O$, HCl and $NH_3$ (*31*). Meanwhile, novograblenovite was also found as a sublimate around vents of hot gases in a few burning coal-dumps, where the crystallization temperature is slightly above 100 °C (*32*). Notably, the latter samples without a basaltic origin is free from K and has a simpler formula $(NH_4)MgCl_3·6H_2O$ (*32*). Compared to the known terrestrial mineral, ULM-1 exhibits a very complicated chemical composition, $[(NH_4)_{0.87}Na_{0.009}K_{0.021}Cs_{0.012}][Mg_{0.97}Ca_{0.023}Al_{0.007}]Cl_3·6H_2O$. Locally, sulfur (S) and phosphorus (P) are also detected in certain parts of the ULM-1 crystal (Supplementary Fig. 3). The heterogeneous distribution of calcium (Ca), aluminum (Al), silicon (Si), sulfur (S), phosphorus (P), and various alkali metals in the ULM-1 crystal aligns with the characteristics of a natural mineral formed from a rock-vapor reaction. Furthermore, isotopic fractionation occurring during the degassing of a hydrous melt would result in elevated $\delta^{37}Cl$ in the vapor phase (*35*). This phenomenon, observed in HCl-bearing terrestrial volcanic gases, aligns with both the presence of $H_2O$ and HCl vapor necessary for the formation of ULM-1 and the observed high $\delta^{37}Cl$ value. The unique isotope and chemical composition of ULM-1, as compared to terrestrial samples, strongly suggests that the possibility of laboratory contamination is highly unlikely.

The oxyhydration of lunar soil in air atmosphere is well-known since the compound FeOOH, arising from the water vapor contamination of indigenous $FeCl_2$, was found in the form of "rust" in lunar rocks from Apollo mission (*43*). However, the possibility of air contact that leads to the formation of water rich ULM-1 crystal is highly unlikely. The formation of ULM-1 can essentially be described by the following chemical equations,

$$6H_2O\,(g) + NH_3\,(g) + HCl\,(g) + MgCl_2\,(s) = (NH_4)MgCl_3·6H_2O\,(s), \quad \text{(eq. 1)}$$

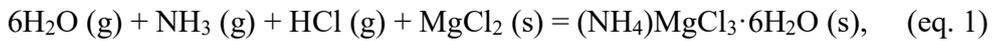

Or

$$6H_2O\,(l) + NH_4Cl\,(s) + MgCl_2\,(s) = (NH_4)MgCl_3·6H_2O\,(s), \quad \text{(eq. 2)}$$

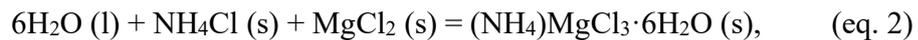

In either case, the formation of ULM-1 needs abundant ammonia species ($NH_3$ or $NH_4Cl$), which is absent in air. The high temperature above 100 °C required for vapor phase reaction (eq. 1) or the aqueous environment (eq. 2) required for the crystallization of ULM-1 further help to rule out the air contact as the source (*31,32*).

The formation conditions of ULM-1 in hot vapor phases raise the possibility of rocket exhaust as the source. However, several factors make the rocket exhaust scenario highly unlikely. Firstly, in-situ spectral analyses conducted by the Chang'E-5 lander on the lunar surface reveal hydroxyl contents within the lower range of lunar hydration features predicted by remote sensing (*25*), with no noticeable influence from rocket exhaust. Additionally, the highest water content observed at the landing site is found in a CE5-Rock with a potential basaltic origin, rather than the lunar regolith, which would be more susceptible to rocket exhaust (*25*). Secondly, the formation of ULM-1 crystal on the lunar surface requires the reaction between hot, magnesium-rich basalt and eruptive gas exhalations enriched in HCl and $NH_3$. However, both eruptive HCl and $NH_3$ are extremely scarce in rocket exhaust plumes, regardless of the type of fuel used in the CE5 main rocket or the lander, such as liquid $H_2/O_2$ and $UDMH/N_2O_4$. Lastly, the exceptionally high Cl isotopic composition of ULM-1 cannot be explained by contamination from terrestrial-sourced rocket exhaust.

Having shown that ULM-1 could have an indigenous lunar origin, it's important to analyze the stability of this hydrate crystal under the harsh lunar environment around the CE5 landing site. It's generally accepted that in the permanently shadowed craters, molecular water could be preserved in the form of ice at low enough temperatures. However, in the sunlit moon, where the lunar condition has become too harsh for ice to survive, the actual chemical form of lunar water ($H_2O$) is still unknown. CE5 spaceship landed at a relatively high latitude 43.058°, where the lunar surface bolometric temperature is not exceeding 80 °C according to the Diviner Dataset (*44*). Moreover, the thermal conductivity of lunar regolith is very low, so that the solar radiation only heats up the topmost grains of a few mm thickness (Supplementary Fig. 7). The direct on-site measurement of lunar surface temperature indicates a 50 °C drop compared to Diviner bolometric temperatures (*45*). For the CE-5 shaded surficial regolith, the highest temperature is expected to be below 30 °C, decreasing to approximately -70 °C for drilled deeper regolith (*45*). Compared to water ice, the decomposition temperature of novograblenovite is reported to be 90 ~ 150 °C higher (*32,46*), and is stable in vacuum below 80 °C (Supplementary Fig. 8). The increased thermally stability of this hydrate should be crucial to protect the lunar water ($H_2O$) from the harsh lunar environment for billions of years. The existence of lunar hydrated salt can naturally explain the water molecule signal recently detected in the sunlit moon (*21*), as well as the enhanced water ($OH^-$ or $H_2O$) signals in some pyroclastic deposits (*47*).

Although the returned lunar basalt is generally highly depleted in water (*18*), the formation condition of this lunar hexahydrate suggests during eruption, some lunar fumaroles have exsolved considerable amount of water and hydrogen vapors, along with other H-bearing species including $NH_3$ and HCl. We performed the thermodynamic calculations of equilibrium constants in a temperature range from 50°C to 150°C (detailed in Methods and Supplementary materials), allowing us to constraint the fugacity of gaseous water $f(H_2O)$ in the lunar volcanic fume. On the earth, novograblenovite was found to crystallize from vapors at temperatures slightly higher than 100°C (*32*). We note the maximum sum of the fugacity of $NH_3$ and HCl is further thermodynamically constrained by the chemical reaction:

$$NH_3\ (g) + HCl\ (g) = NH_4Cl\ (s) \qquad (eq.3)$$

Thus the temperature dependent low-limit $f(H_2O)$ to form ULM-1 can be obtained and delineated by the red curve in Fig. 4. Below this equilibrium line, the hexahydrate becomes unstable and gives way to a known dihydrate - $(NH_4)MgCl_3·2H_2O$ (*46*). The upper-limit of $f(H_2O)$ is set by the equilibrium between the liquid water and its saturated vapor (blue line). In this temperature range, the low-limit of $f(H_2O)$ of lunar volcanic gas is constrained to be 3.5%~16.6% of the saturated water vapor. With reference to the formation temperature of terrestrial novograblenovite (>100°C) (*32*), the $f(H_2O)$ of the volcanic gas should be greater than 7.7% atm. As a comparison, this low-limit is around 3 times lower than the $f(H_2O)$ of the driest Earth volcano – Lengai volcano (>24% atm) (*48*). It worth noting that in Tolbacik volcano, where the earth mineral is located, the measured $f(H_2O)$ is higher than 92% atm (pentagon) (*3*), one lg unit higher than the thermodynamic equilibrium value. Moreover, by taking $H_2$ degassing into account (*19, 49*), the estimated $f(H_2O + H_2)$ values (purple doted line) become comparable to Lengai volcano on earth (detailed in Supplementary materials). Considering the potentially distinct volatile contents of pyroclastic eruptions and mare basalts, our results depict a specific phase of lunar volcanic eruption during the crystallization of ULM-1.

The existence of lunar volcanic gases rich in $H_2O$ and $H_2$ vapors also has implications for the origin of the volatiles detected in permanently shadowed craters. The remote sensing missions have shown that in addition to $H_2O$, there is significant amount of solid $H_2$ and $NH_3$ in the lunar polar regions (*16,50*), which is within the lunar volcanic gas inventory revealed here. It is conceivable that some of these volatiles, including water ($H_2O$), could have originated from magmatic degassing during eruption of ancient lunar volcanos. After more than half a century of lunar sampling missions and laboratory research, the presence of molecular water ($H_2O$) is discovered in lunar soil, in the form of a well-crystallized hydrate. The structure and composition of this hydrate closely resemble that of a rare terrestrial fumarole mineral - novograblenovite, suggesting that the lunar mineral $[(NH_4)_{0.87}Na_{0.009}K_{0.021}Cs_{0.012}][Mg_{0.97}Ca_{0.023}Al_{0.007}]Cl_3·6H_2O$ could be formed through reactions involving basaltic lava and lunar volcanic gases rich in $H_2O$, HCl, and $NH_3$. The low formation temperature of this mineral can also be achieved through diurnal heating or impact gardening on the moon. While lunar water has typically been attributed to sources such as solar wind, comets, and meteorites, the remarkably high $\delta^{37}Cl$ value of this hydrate suggests that it is more likely indigenous to the moon. The enhanced thermal stability of this hexahydrate, compared to water-ice, coupled with its high latitude location, may contribute to the preservation of molecular water on the sunlit moon. This hydrate provides critical constraints on the lowest water fugacity $f(H_2O)$ of lunar volcanic gases during its crystallization, with the resulting $f(H_2O+H_2)$ falling within the range observed in some of the driest volcanoes on Earth.

## Methods

### Energy-dispersive X-ray spectrometer (EDS).

Unlike known lunar mineral samples, ULM-1 evaporate quickly under electron beams of EDS and especially EPMA. The concentrations of main elements of ULM-1 were acquired using a Phenom pro XL microscope equipped with an electron microprobe analyzer for the elemental analysis in the EDS mode (15 kV acceleration voltage; 5 nA beam current). The data acquisition rate was ~$10^3$ counts/s, with 20–25% dead time and 200 s live time. Element concentrations were measured using the Kα lines for N, O, K, Mg and Cl. X-ray intensities were converted to wt% by the XPP quantitative analysis software of Oxford Instruments (UK). The standards employed for quantitative analysis were bischofite ($MgCl_2·6H_2O$, for elements Mg, O and Cl) and sal-ammoniac ($NH_4Cl$, for Cl and N). The concentrations of trace elements of ULM-1 were acquired using a Hitachi S4800 microscope equipped with an electron microprobe analyzer EMAX in the EDS mode (15 kV acceleration voltage; 10 nA beam current). The data acquisition rate was ~$10^3$ counts/s, with 20–25% dead time and 1200 s live time. The concentrations of elements exceeding 0.1 wt. % can be analyzed based on this method. The results of EDS data are present in Supplementary Fig. 1-3, Supplementary Table 1.

### Electron probe microanalysis (EPMA).

The composition of ULM-1 was analyzed using a Shimadzu 8050G electron probe microanalyser. The ULM-1 sample prepared for EPMA analysis is 40 × 50 × 20 μm in size, and the material was carbon coated and prepared as flat sections before insertion into the vacuum. While the spectrum weight of O, N and Cl volatiles decreased substantially relative to Mg under strong EPMA beams, the concentrations of more stable minor elements of K, Cs, Na, Ca, Al and Si were acquired with low detection limits down to 100 ppm. An acceleration voltage of 15 kV and beam current of 50 nA were used for all analyses, with a spot size of 10 μm and integration time of 600 s. Data were processed with the ZAF matrix correction using EPMA-Browser software. Forsterite were used as the standards for Mg. The standards used for the other elements analyzed were sodium chloride (chlorine (Cl)), silicon dioxide (oxygen (O)), boron nitride (nitrogen (N)), potassium feldspar (potassium (K)), sodium feldspar (sodium (Na)), pollucite (cesium (Cs)), elemental aluminum (aluminum (Al)), elemental silicon (silicon (Si)) and calcium fluoride (calcium (Ca)). The analysis results for ULM-1 are listed in Supplementary Table 1.

**Raman measurements.**

Raman scattering experiments have been performed in a quasi-backscattering geometry with the excitation line λ = 532 nm of a solid-state laser (P=2 mW). The laser beam was focused to a 3-μm-diameter spot on the surface of the ULM-1 crystal. The spectra were recorded using a WITec α-300R and JY Horiba HR800 spectrometer.

**Infrared spectroscopy.**

The infrared (IR) spectrum of ULM-1 was obtained using a SpectrumOne FTIR spectrometer coupled with an AutoImage microscope with a spectral resolution of 1 or 2 cm$^{-1}$. The sample was placed on a gold mirror, and the spectrum obtained is a superposition of transmission and reflectance signals with strong predominance of the former. Rectangular apertures matching crystal dimensions (i.e. 50 × 50 μm) were employed.

**X-ray crystallography.**

X-ray diffraction intensities of reflections were collected for ULM-1 single crystal on a Bruker D8 Venture single-crystal diffractometer equipped with a micro-focus X-ray tube (multilayer mirror monochromatized Mo K$_α$ radiation) at room temperature. Data collection, cell refinement, and data reduction were carried out in the Bruker APEX4 program. All calculations were performed with SHELX programs in the framework of an Olex2 software package (*51*). The ULM-1 mineral crystallizes in the monoclinic space group C2/c, with lattice parameters a=9.3222(10) Å, b=9.5731(10) Å, c= 13.3328(12) Å and β=90.101(5) °. The crystal structure was successfully solved via intrinsic phasing method by SHELXT and refined with SHELXL against the $F^2$ data to the final R factor of 0.0246 (for 1372 unique reflections with $F_O > 4$ sig ($F_O$)) with anisotropic displacement parameters for all nonhydrogen atoms. Refinements of the occupancies of Mg and N sites were determined based on the results of elemental composition determination by EPMA. Ammonium ions were imposed by rigid body geometry constraints, further releasing the N atom from a symmetry element allowed a flexible split of N inside a cage of 6 Cl⁻ vertices and decreased the anisotropic displacement parameters giving a better fit to the collected X-ray intensities. The results of crystallography data are present in Supplementary Table 2-5.

**Measurement of chlorine isotopes in terrestrial samples using pulsed thermal ionization mass spectrometry**

1. Growth of the terrestrial material as isotopic standard

The chemicals $NH_4Cl$ and $MgCl_2·6H_2O$ were purchased from Sinopharm Reagent Corporation and were of analytical grade with a purity greater than 99.0%. They were used as received without any additional purification. Deionized water was used to prepare solutions and for subsequent chemical analysis. To prepare $NH_4MgCl_3·6H_2O$ single crystals, a mixture of $NH_4Cl$ and $MgCl_2·6H_2O$ in a mole ratio of 1:1 was added to a beaker and stirred in distilled water until fully dissolved. The solution was then subjected to slow evaporation, resulting in the formation of colorless crystals of $NH_4MgCl_3·6H_2O$ *(52)*. The crystals obtained in this manner are used as standard material for isotopic analysis and thermodynamic analysis.

2. Sample treatment of the terrestrial material for isotropic measurement

All samples were prepared using a two-step resin method. First, regenerated H-cation exchange resin (Dowex 50 W × 8, 200-400 mesh) was packed into a polyethylene-accompanied ion-exchange column with a diameter of 0.4 cm and a resin height of 2 cm. Subsequently, regenerated Cs-cation exchange resin was packed into another polyethylene-accompanied ion-exchange column with a diameter of 0.4 cm and a resin height of 1.6 cm. The liquid samples were passed through the H-cation exchange resin column, where all the Cl in the samples transformed into HCl solution. Then, the liquid passed through the Cs-cation exchange resin column, and the CsCl solution collected was used for mass spectrometry analysis. The pH of the solution was neutral.

3. Pulsed thermal ionization mass spectrometry measurement procedure

Chlorine isotopes of terrestrial material were analyzed using the Pulsed Thermal Ionization Mass Spectrometry (P-TIMS) method in a Triton mass spectrometer *(53)*. The tantalum filament was heated with a current of 2.5 A for 1 hour in a vacuum system and coated with 2.5 μL of graphite slurry (80% ethanol mixed with 80 μg of graphite) at the center of the filament. The test solution, consisting of CsCl as the source of chlorine, was loaded onto the filament and dried using a current of 1 A for 1.5 minutes. The samples were then placed in the mass spectrometer, and the measurement was initiated once the ion source was evacuated to a pressure lower than $2.5 \times 10^{-7}$ mbar. The intensity of the $Cs_2Cl^+$ ion was adjusted to $4 \times 10^{-12}$ A by controlling the filament current. Data were simultaneously collected on Faraday cup C and H1 by capturing the ion flow with mass numbers of 301 ($^{133}Cs_2{}^{35}Cl^+$) and 303 ($^{133}Cs_2{}^{37}Cl^+$).

The $\delta^{37}Cl$ values were calculated using the following equation:

$$\delta^{37}Cl\ (‰) = [(^{37}Cl/^{35}Cl)_{Sample}/(^{37}Cl/^{35}Cl)_{Standard} - 1] \times 1000$$

The average value of the standard material ISL 354 NaCl was obtained from the results of three repeated determinations under the same conditions. The measured value of the international standard ISL 354 for chlorine isotopes is 0.319062±0.00005. The measured $\delta^{37}Cl$ value of the terrestrial $NH_4MgCl_3 \cdot 6H_2O$ single crystals is 0.08 ‰ (relative to SMOC).

**Measurement of Cl Isotope in ULM-1 Sample using nano secondary ion mass spectrometer**

The in-situ isotopic measurements of chlorine were performed in the NanoSIMS Lab at School of Earth system science (SESS), Tianjin University. The CAMECA NanoSIMS 50L secondary ion mass spectrometer is used for Cl isotope measurements of ULM-1, the terrestrial $NH_4MgCl_3 \cdot 6H_2O$ crystals with a determined $\delta^{37}Cl$ (0.08 ‰ relative to SMOC) is used to calibrate the matrix effects and instrument mass fractionation (IMF) of novograblenovite. To prepare the NanoSIMS specimen, flat pieces of terrestrial $NH_4MgCl_3 \cdot 6H_2O$ (160×120×15 μm) and ULM-1 (80×60×15 μm) crystals were fixed on a 10 mm diameter Si disk with thin films of carbon paste. The analytical spatial resolution was about ~100 nm using a $Cs^+$ primary beam with an intensity of 1 pA and an accelerating voltage of ~16 kV. Each analysis surface area with a raster size of 5 μm was divided into 64 ×64 pixels, with a counting time of 0.132 ms per pixel. The number of cycles was set at 300. Six minutes of pre-sputtering using a 300 pA was set-up before analysis on a larger area (30×30μm$^2$) than the analyzed surface to remove any contamination from the surface and establish sputter equilibrium. An electron flood gun was used for charge compensation. Secondary negative ions of $^{16}$O, $^{35}$Cl, $^{37}$Cl were imaged by scanning ion imaging, with a MRP of 6000.

For Cl isotope measurements, the instrumental mass fractionation (IMF) factor, α, based on analyses of terrestrial crystal standard is 1.0212 ± 0.0008(2SD). The measured $^{37}$Cl/$^{35}$Cl ratios are corrected for the IMF. And the quasi-simultaneous arrival (QSA) effect was calibrated using the relation *(54,55)*.

$$N_{true} = N_{meas} \times (1 + \beta \times K)$$

where $N_{meas}$ is the measured ion count, $N_{true}$ is the true ion count, K is the average number of secondary ions ejected per primary ion, expressed as $I_{secondary}/I_{primary}$ (the intensity ratio of the secondary ions to the primary ions that hit the sample surface, i.e. FCo), and β is the QSA coefficient. The results of Cl isotopes data are present in Supplementary Table 6.

**Lunar and terrestrial materials.**

The lunar sample, designation CE5C0400YJFM00507 was provided by the China National Space Administration (CNSA) under a materials transfer agreement a half-year loan (with an extension for another half-year), after which they will be returned to CNSA. To obtain large enough samples (>1 mg) to measure the specific heat data of (NH$_4$)MgCl$_3$·6H$_2$O (S6) and (NH$_4$)MgCl$_3$·2H$_2$O (S2) for thermodynamic analysis, we synthesized high purity crystals of S6 based on a modified solution method described above and in Ref. 51. Powder samples of S2 is obtained by heating the S6 in air at 140°C.

**Thermodynamic analysis of the fugacity of water vapor.**

**A: (NH$_4$)MgCl$_3$·6H$_2$O (abbreviated as S6 here below)：**

The heat capacity of S6 single crystal is measured in Quantum Design physical property measurement system.

Fitting the heat capacity data.

According to Debye model, the heat capacity

$$C_V(T) = 9NR(T/\theta_D)^3 \int_0^{\theta_D} \frac{x^4 e^x}{(e^x - 1)^2} dx$$

Here $\theta_D$ is the Debye temperature, $R$ the ideal gas constant, N=27 because there are 27 moles of atoms in each mole of S6 compound.

According to Einstein model, the heat capacity

$$C_V(T) = 3NR\left(\frac{\theta_E}{T}\right)^2 \frac{e^{\frac{\theta_E}{T}}}{(e^{\frac{\theta_E}{T}} - 1)^2}$$

Here $\theta_E$ is the Einstein temperature.

Debye model can usually well describe the heat capacity of a simple compound. For crystals with weakly bonded structure unites, such as the S6 with [Mg(H$_2$O)$_6$] and [NH$_4$] blocks, additional Einstein model should be used to describe the local vibrations of such unites. To fit the heat capacity of S6, one Debye model and two independent Einstein models are used.

By fitting the above equations to the measured data (Supplementary Table 7 & 8), the function $C_V(T)$ is given, as shown by the blue line in Supplementary Figure 9.

Under constant pressure:

$$C_P(T) = C_V(T) + P\, dV/dT = C_V(T) + PV(dV/VdT)$$

Here V is the mole volume, $dV/VdT$ is thermal expansion coefficient.

At standard pressure, $P^o = 1.01 \times 10^5\, Pa$, V(S6) = 170.7×10$^{-6}$ m³/mol. For solid substances, $dV/VdT$ is typically smaller than 10$^{-4}$, so that $PdV/VdT$ is on the order of 10$^{-3}$ J/(mol K), which can be safely ignored considering the Cv is generally on the order of 10$^5$ J/(mol K). Here below, the fitted $C_V(T)$ is used as $C_P(T)$ for S6 without further correction.

The definition of entropy and enthalpy are as follows:

$$S^o(T) = \int_0^T \frac{C_P(t)}{t} dt \quad \text{(eq. 4)}$$

$$H^o(T) = H^o(0) + \int_0^T C_P(t)\, dt \quad \text{(eq. 5)}$$

The entropy of the S6 crystal thus obtained is shown in Supplementary Figure 10.

The formation of ULM-1 can be described by the following chemical equation during the volcanic erupt,

$$6H_2O \text{ (g)} + NH_3 \text{ (g)} + HCl \text{ (g)} + MgCl_2 \text{ (s)} = (NH_4)MgCl_3 \cdot 6H_2O \text{ (s)} \quad \text{(eq. 6)}$$

The entropy change in the formation of ULM-1 is then:

$$\Delta S^o = S^o((NH_4)MgCl_3 \cdot 6H_2O) - 6 \times S^o(H_2O) - S^o(NH_3) - S^o(HCl) - S^o(MgCl_2)$$

The entropy of $NH_3$ (g), HCl (g), $H_2O$ (g) and $MgCl_2$ (s) are taken from the JANAF-NIST thermal-dynamic data base, and the resulting entropies are shown in Supplementary Figure 11.

The enthalpy change in the formation of ULM-1 is then:

$$\Delta H^o = H((NH_4)MgCl_3 \cdot 6H_2O)^{solid} - 6 \times H(H_2O)^{gas} - H(NH_3)^{gas} - H(HCl)^{gas} - H(MgCl_2)^{solid}$$

Considering the above enthalpy change $\Delta H^o$ is mostly contributed by the solid to gas transition, the $\Delta H^o$ can be reformed to:

$$\Delta H^o = \{H((NH_4)MgCl_3 \cdot 6H_2O)^{solid} - [6 \times H(H_2O)^{solid} + H(NH_4Cl)^{solid} + H(MgCl_2)^{solid}]\}$$
$$+ \{[6 \times H(H_2O)^{solid} + H(NH_4Cl)^{solid} + H(MgCl_2)^{solid}]$$
$$- [6 \times H(H_2O)^{gas} + H(NH_3)^{gas} + H(HCl)^{gas} + H(MgCl_2)^{solid}]\}$$

Where we define the enthalpy change between a solid phase transition:

$$\Delta H_1 = \{H((NH_4)MgCl_3 \cdot 6H_2O)^{solid} - [6 \times H(H_2O)^{solid} + H(NH_4Cl)^{solid} + H(MgCl_2)^{solid}]\}$$

And the enthalpy difference between solid and gaseous reagent is:

$$\Delta H_2 = \{[6 \times H(H_2O)^{solid} + H(NH_4Cl)^{solid} + H(MgCl_2)^{solid}]$$
$$- [6 \times H(H_2O)^{gas} + H(NH_3)^{gas} + H(HCl)^{gas} + H(MgCl_2)^{solid}]\}$$
$$= \{6 \times [H(H_2O)^{solid} - H(H_2O)^{gas}]$$
$$+ [H(NH_4Cl)^{solid} - H(NH_3)^{gas} - H(HCl)^{gas}]\}$$

At 0K, the $\Delta H_2$ obtained from the JANAF-NIST thermal-dynamic database is:

$$\Delta H_2 = -463.179 \text{ KJ/mol}$$

Meanwhile, the $\Delta H_1$ is much smaller and its value at 0K can be obtained based on the ab-initial quantum chemistry calculations:

$$\Delta H_1 = -54.499 \text{ KJ/mol}$$

So that the $\Delta H^o(0)$ in equation 5 is obtained as:

$$\Delta H^o(0) = \Delta H_1 + \Delta H_2 = -517.679 \text{ KJ/mol}$$

At finite temperatures, the enthalpy change of equation (6) is determined by the specific heat difference between the product and the reagent:

$$\Delta H^o(T) = \Delta H^o(0) + \int_0^T \Delta C_P(t)\, dt$$

$$= \Delta H^o(0) + \int_0^T C_P^{((NH_4)MgCl_3 \cdot 6H_2O)^{solid}}(t)\, dt$$

$$- \int_0^T \left[ 6 \times C_P^{(H_2O)^{gas}}(t) + C_P^{(HCl)^{gas}}(t) + C_P^{(NH_3)^{gas}}(t) + C_P^{(MgCl_2)^{solid}}(t) \right] dt$$

$$= \Delta H^o(0) + \int_0^T C_P^{((NH_4)MgCl_3 \cdot 6H_2O)^{solid}}(t)\, dt$$

$$- \left\{ 6 \times \left[ H^{(H_2O)^{gas}}(T) - H^{(H_2O)^{gas}}(0) \right] + \left[ H^{(HCl)^{gas}}(T) - H^{(HCl)^{gas}}(0) \right] \right.$$

$$\left. + \left[ H^{(NH_3)^{gas}}(T) - H^{(NH_3)^{gas}}(0) \right] + \left[ H^{(MgCl_2)^{solid}}(T) - H^{(MgCl_2)^{solid}}(0) \right] \right\}$$

Based on the $\Delta H$ data taken from the JANAF-NIST thermal-dynamic database and the fitted specific heat of ULM-1, the $\Delta H^o(T)$ is shown in Supplementary Figure 12.

The reaction Gibbs free energy of the chemical equation (7) can be obtained based on the reaction entropy and reaction enthalpy:

$$\Delta G^o = \Delta H^o - T\Delta S^o$$

which is shown in Supplementary Figure 13.

According to chemical reaction isotherm,

$$\ln K = -\Delta G_r^o(T)/RT$$

Here $K$ is the equilibrium constant, $R$ the ideal gas constant, the result is shown in Supplementary Figure 14.

Therefore, the equilibrium constant for the reaction equation (6) of the hexahydrate can be obtained as:

$$lgK_6 = \frac{-\Delta G_r^o(T)}{ln10 \times RT}$$

The equilibrium constant K of reaction equation (6) is:

$$K_6 = \frac{1}{\left[\frac{f(H_2O)}{P^o}\right]^6 \frac{f(HCl)}{P^o} \frac{f(NH_3)}{P^o}}$$

Here $P^o$ is the standard pressure, $f(H_2O)$, $f(HCl)$ and $f(NH_3)$ are the fugacity of water, HCl and NH$_3$ vapors, respectively.

Therefore:

$$lgK_6 = \frac{-\Delta G_6^o(T)}{ln10 \times RT} = -\left\{6lg\left[\frac{f(H_2O)}{P^o}\right] + lg\frac{f(HCl)}{P^o} + lg\frac{f(NH_3)}{P^o}\right\} \quad (eq.\ 7)$$

The sum of $f(HCl)$ and $f(NH_3)$ in volcanic gases is thermodynamically constraint by a well-known chemical reaction:

$$NH_3\ (g) + HCl\ (g) = NH_4Cl\ (s) \quad (eq.\ 8)$$

Similarly, we have:

$$lgK_{NH4Cl} = \frac{-\Delta G_{NH4Cl}^o(T)}{ln10 \times RT} = -\left\{lg\frac{f(HCl)}{P^o} + lg\frac{f(NH_3)}{P^o}\right\}$$

Which set a thermodynamic maximum value of $\left\{lg\frac{f(HCl)}{P^o} + lg\frac{f(NH_3)}{P^o}\right\}$

$$\left\{lg\frac{f(HCl)}{P^o} + lg\frac{f(NH_3)}{P^o}\right\}^{max} = \frac{\Delta G_{NH4Cl}^o(T)}{ln10 \times RT} \quad (eq.\ 9)$$

which is shown in Supplementary Figure 15.

Combing equation (7) and (9), we have:

$$6lg\left[\frac{f(H_2O)}{P^o}\right] = \frac{\Delta G_6^o(T)}{ln10 \times RT} - \left\{lg\frac{f(HCl)}{P^o} + lg\frac{f(NH_3)}{P^o}\right\}$$

$$> \frac{\Delta G_6^o(T)}{ln10 \times RT} - \left\{lg\frac{f(HCl)}{P^o} + lg\frac{f(NH_3)}{P^o}\right\}^{max} = \frac{\Delta G_6^o(T) - \Delta G_{NH4Cl}^o(T)}{ln10 \times RT}$$

So that we get the thermal-dynamical lower limit of $f(H_2O)$:

$$\left\{lg\left[\frac{f(H_2O)}{P^o}\right]\right\}^{min} = \frac{\Delta G_6^o(T) - \Delta G_{NH4Cl}^o(T)}{ln10 \times 6RT} \quad \text{(eq. 10)}$$

Wherein the $\Delta G_6^o(T)$ is shown in Supplementary Figure 13, the $\Delta G_{NH4C}^o(T)$ is obtained from the JANAF-NIST thermal-dynamic database based on the formation Gibbs energy of NH$_4$Cl (s), NH$_3$ (g) and HCl (g). The resulting $\left\{lg\left[\frac{f(H_2O)}{P^o}\right]\right\}^{min}$ for $(NH_4)MgCl_3 \cdot 6H_2O$ is shown in Figure 4.

**B: (NH$_4$)MgCl$_3$·2H$_2$O (abbreviated as S2 here below):**

Similar to the thermodynamic analysis shown above for the hexahydrate $(NH_4)MgCl_3 \cdot 6H_2O$, the heat capacity, reaction entropy, reaction enthalpy, reaction Gibbs free energy, equilibrium constant and the minimum fugacity of water can also be obtained for the dehydrate $(NH_4)MgCl_3 \cdot 2H_2O$ following the reaction equation (11):

$$2H_2O \text{ (g)} + NH_3 \text{ (g)} + HCl \text{ (g)} + MgCl_2 \text{ (s)} = (NH_4)MgCl_3 \cdot 2H_2O \text{ (s)} \quad \text{(eq. 11)}$$

The heat capacity of S2 powder sample is measured in Quantum Design physical property measurement system. By fitting the Debye&Einstein equations to the measured data (Supplementary Table 7 & 9), the function C$_V$(T) is given, as shown by the blue line in Supplementary Figure 16.

The entropy change in the formation of S2 is then:

$$\Delta S^o = S^o\big((NH_4)MgCl_3 \cdot 6H_2O\big) - 2 \times S^o(H_2O) - S^o(NH_3) - S^o(HCl) - S^o(MgCl_2)$$

the result of which is shown in Supplementary Figure 17.

The entropy of NH$_3$ (g), HCl (g), H$_2$O (g) and MgCl$_2$ (s) are taken from the JANAF-NIST thermal-dynamic data base, and the resulting entropies are shown in Supplementary Figure 18.

The enthalpy change in the formation of S2 is:

$$\Delta H_2 = H\big((NH_4)MgCl_3 \cdot 2H_2O\big)^{solid} - 2 \times H(H_2O)^{gas} - H(NH_3)^{gas} - H(HCl)^{gas} - H(MgCl_2)^{solid}$$

Considering the enthalpy change ΔH during the decomposition of S6 to S2 is known, the reaction enthalpy $\Delta H_2$ can be obtained based on $\Delta H_6$ and the heat capacity of S6, S2 and H$_2$O vapors:

$$\Delta H_6 = H\big((NH_4)MgCl_3 \cdot 6H_2O\big)^{solid} - 6 \times H(H_2O)^{gas} - H(NH_3)^{gas} - H(HCl)^{gas}$$
$$- H(MgCl_2)^{solid}$$

$$\Delta H_6 - \Delta H_2 = H\big((NH_4)MgCl_3 \cdot 6H_2O\big)^{solid} - H\big((NH_4)MgCl_3 \cdot 2H_2O\big)^{solid} - 4 \times H(H_2O)^{gas}$$

The right side of the above equation is the enthalpy change during the decomposition of S6 to S2 and water vapor, which is defined by:

$$\Delta H^D(T) = \Delta H^D(T_D) + \int_{T_D}^{T} \Delta C_P(t)\, dt$$

$$= \Delta H^D(T_D) + \int_{T_D}^{T} C_P^{((NH_4)MgCl_3 \cdot 6H_2O)^{solid}}(t) - C_P^{((NH_4)MgCl_3 \cdot 2H_2O)^{solid}}(t) - 4 C_P^{(H_2O)^{gas}}(t)\, dt$$

So that:

$$\Delta H_2(t) = \Delta H_6(t)$$
$$- \Bigg\{ \Delta H^D(T_D)$$
$$+ \int_{T_D}^{T} C_P^{((NH_4)MgCl_3 \cdot 6H_2O)^{solid}}(t) - C_P^{((NH_4)MgCl_3 \cdot 2H_2O)^{solid}}(t)$$
$$- 4 C_P^{(H_2O)^{gas}}(t)\, dt \Bigg\}$$

Where the $\Delta H^D(T_D)$ is reported to be 240±20 KJ/mol, $T_D$ is reported to be 428 K *(56)*, the $\Delta H_6(t)$ is taken from Supplementary Figure 12, the heat capacity of S6, S2 and water (gas) are taken from Supplementary Figure 9&16 and the JANAF-NIST thermal-dynamic database.

The $\Delta H_2(t)$ can thus be obtained and is shown in Supplementary Figure 19.

The reaction Gibbs free energy of the chemical equation (11) can be obtained based on the reaction entropy and reaction enthalpy:

$$\Delta G^o = \Delta H^o - T\Delta S^o$$

The result of which is shown in Supplementary Figure 20.

The equilibrium constant K and water fugacity $f(H_2O)$ for S2 can then be obtained based on a similar thermodynamic analysis of the S6:

$$\ln K = -\Delta G_r^o(T)/RT$$

Here $K$ is the equilibrium constant, $R$ the ideal gas constant.

Therefore, the equilibrium constant for the reaction equation (11) of the dehydrate can be obtained as:

$$lgK_2 = \frac{-\Delta G_r^o(T)}{ln10 \times RT}$$

which is shown in Supplementary Figure 21.

The equilibrium constant K of reaction equation (6) is:

$$K_2 = \frac{1}{\left[\frac{f(H_2O)}{P^o}\right]^2 \frac{f(HCl)}{P^o} \frac{f(NH_3)}{P^o}}$$

Here $P^o$ is the standard pressure, $f(H_2O)$, $f(HCl)$ and $f(NH_3)$ are the fugacity of water, HCl and NH$_3$ vapors, respectively.

Therefore:

$$lgK_2 = \frac{-\Delta G_2^o(T)}{ln10 \times RT} = -\left\{2lg\left[\frac{f(H_2O)}{P^o}\right] + lg\frac{f(HCl)}{P^o} + lg\frac{f(NH_3)}{P^o}\right\} \quad (12)$$

Incorporating the thermodynamic maximum value of $\left\{lg\frac{f(HCl)}{P^o} + lg\frac{f(NH_3)}{P^o}\right\}$ based on equation (9) with (12), we have:

$$2lg\left[\frac{f(H_2O)}{P^o}\right] = \frac{\Delta G_2^o(T)}{ln10 \times RT} - \left\{lg\frac{f(HCl)}{P^o} + lg\frac{f(NH_3)}{P^o}\right\}$$

$$> \frac{\Delta G_2^o(T)}{ln10 \times RT} - \left\{lg\frac{f(HCl)}{P^o} + lg\frac{f(NH_3)}{P^o}\right\}^{max} = \frac{\Delta G_2^o(T) - \Delta G_{NH4Cl}^o(T)}{ln10 \times RT}$$

So that we get the thermal-dynamical lower limit of $f(H_2O)$:

$$\left\{lg\left[\frac{f(H_2O)}{P^o}\right]\right\}^{min} = \frac{\Delta G_2^o(T) - \Delta G_{NH4Cl}^o(T)}{ln10 \times 2RT} \quad (eq.\ 13)$$

Wherein the $\Delta G_2^o(T)$ is shown in Supplementary Figure 20, the $\Delta G_{NH4Cl}^o(T)$ is obtained from the JANAF-NIST thermal-dynamic database, the resulting $\left\{lg\left[\frac{f(H_2O)}{P^o}\right]\right\}^{min}$ for $(NH_4)MgCl_3 \cdot 2H_2O$ is shown in Figure 4.

**C: H₂ and H₂O gases in magma**

The stability of water in lunar magma is governed by the reaction

$$H_2O = H_2 + 1/2 O_2 \quad (11)$$

And is therefore a function of both the oxygen and hydrogen fugacity of the system. Thermodynamically, reaction (11) is governed by the equation

$$\ln \frac{\left[\frac{f(H_2O)}{P^o}\right]}{\frac{f(H_2)}{P^o}\left[\frac{f(O_2)}{P^o}\right]^{1/2}} = -\Delta G_r^o(T)/RT$$

there are no reliable mineral buffers for the hydrogen fugacity of lunar materials, but the $f(O_2)$ of the Moon can be used to estimate $f(H_2)/f(H_2O)$ ratios.

The $f(O_2)$ of the lunar mineral assemblages and lunar igneous rocks has been intensively investigated and is found clustered around the Fe-ilmenite-ulvöspinel (IIU) buffer

$$2Fe + 2FeTiO_3 + O_2 = 2Fe_2TiO_4$$

which is shown in Supplementary Figure 22.

The melting and crystallization relationships in synthetic melts of mare basalt composition have been extensively studied in the laboratory. The temperature at which the molten high-Ti mare basalts began to crystallize, called the liquidus temperature, is generally around 1150°C (18). Meanwhile, when the basalt was heated, they start to melt at around 1050°C, this melting temperature is about 100°C higher than those of terrestrial basalts, chiefly because of the lake of H₂O in returned lunar basalts (18). Based on the $\Delta G_r^o(T)$ values obtained from the JANAF-NIST thermal-dynamic database, The calculated $f(H_2)/f(H_2O)$ ratios at IIU buffer and 1100°C magma temperature is 2.95. In other words, H₂ is the predominant vapor species in the O-H system at the low $f(O_2)$ of the Moon. It is noteworthy that Newcombe et al. have also examined the solubility of H₂ and H₂O in lunar magmas, estimating a very similar f(H₂)/f(H₂O) ratio of around 2.67 (19). Both the results of Newcombe et al. and our thermodynamic calculations are used to estimate the $f(H_2O) + f(H_2)$ values. The resulting $\left\{lg\left[\frac{f(H_2O)+f(H_2)}{P^o}\right]\right\}^{min}$ values constrained by the formation of $(NH_4)MgCl_3 \cdot 6H_2O$ is shown in Figure 4.

## Data availability

The lunar sample, designation CE5C0400YJFM00507 was provided by the China National Space Administration (CNSA) under a materials transfer agreement a half-year loan (with an extension for another half-year), after which they will be returned to CNSA. Readers may request Chang'e-5 samples from CNSA through a standard procedure. All data are available in the main text or the Supplementary Materials. Source data are provided in this paper. The X-ray crystallographic coordinates for structures reported in this study have been deposited at the Cambridge Crystallographic Data Centre (CCDC), under deposition numbers 2166870. These data can be obtained free of charge from The Cambridge Crystallographic Data Centre via www.ccdc.cam.ac.uk/data_request/cif.

## Acknowledgments

The Chang'e-5 lunar sample CE5C0400YJFM00507 (1.5 g) was provided by the China National Space Administration. We thank Ying Li, Qinghua Zhang, Xuefeng Wang, Ke Ma, Qi Li, Junyan Zhou, Tianping Ying from IOP, CAS, Chuanqiang Sun from Tianjin University and Chunlai Li, Bin Liu from National Astronomical Observatories, CAS for their assistance in experiments and useful discussions. This work was supported by the Key Research Program of Chinese Academy of Sciences (Grant Number ZDBS-SSW-JSC007-2, XLC), the Strategic Priority Research Program and Key Research Program of Frontier Sciences of the Chinese Academy of Sciences (Grant No. XDB33010100, XLC), National Natural Science Foundation of China (Grants No. 52272268, S.F.J.) and the Youth Innovation Promotion Association of CAS (Grant Number 2019005, SFJ).

## Author contributions statement

SFJ conducted sample selection, EDX, EPMA, IR, Raman, Isotopic experiments, thermodynamic calculations, analysis the data and wrote the manuscript, MNH conducted sample selection, single out the ULM-1 mineral, determined the crystal structure, ZNG performed thermodynamic calculations, BHY performed Isotopic experiments, YXM performed EDX experiments, LGD performed the EPMA measurements, CC captured the Optical photos, YPS performed Raman experiments, CC performed Raman experiments, CCC performed Raman experiments, KM performed Isotopic experiments, QW performed Isotopic experiments, YQM performed Isotopic experiments, GJG performed Raman experiments, XLC supervised the project, analysis the data and wrote the manuscript.

**Conflict of interests statement**

The authors declare no competing interests.

**Figure legends/captions**

**Figure 1. Photograph and composition of lunar hydrous mineral ULM-1.** (**a**) Chang'e-5 soil samples. (**b**) Photo of ULM-1 single crystal mounted on the top of a cactus thorn. (**c**) Energy-dispersive X-ray spectroscopy (EDX) spectrum, the data (cyan shading) can be fitted by the presence of N, Mg, O and Cl elements (the black line). (**d**) Electron probe microanalysis (EPMA) spectra show the presence of trace elements Cs, Ca, K, Al, Si and Na. (**e**) Raman spectrum of ULM-1. (**f**) Infrared spectrum of ULM-1. The a.u. represents arbitrary units.

**Figure 2. Crystal structure and charge density of lunar hydrate ULM-1.** (**a**) Crystal structure of lunar mineral $(NH_4)MgCl_3 \cdot 6H_2O$, the dotted lines present a network of hydrogen bond interactions around the $H_2O$ molecules and $NH_4^+$ ions. (**b**) The mineral crystalized in a perovskite-like structure. (**c**) Differential charge density analysis around all the oxygen atoms shows clearly the H peaks. (**d**) The atoms in an asymmetric unit, where three independent $H_2O$ molecules in a $[Mg(H_2O)_6]^{2+}$ unit and the $NH_4^+$ motif are interconnected by $Cl^-$.

**Figure 3. Chlorine isotope variability in different terrestrial and extraterrestrial reservoirs.** Chlorine isotope composition of ULM-1 is shown in Cyan pentagram, data are presented as mean values of 5 measurements ± standard deviation (σ). Open circles are liquids and vapors, Seawater: regular pentagon; porewaters: diamond; brines/formation waters: hexagon; volcanic gases: square; Solid circles are solid minerals, natural perchlorates: blue triangle; evaporites: green square; sediments/sedimentary rocks: dark cyan triangle; metasedimentary rocks: purple triangle; altered oceanic crust: orange pentagon; volcanic ashes/lavas: green square; MORB: yellow triangle; OIB: brown diamond. Dashed line is the highest value recorded in terrestrial minerals. Gray band delineates −1.0 to +1.0 ‰. Almost all evaporite and MORB values fall within this gray band

illustrating that the largest Cl terrestrial reservoirs are near 0‰. Samples from the Mars, Chondrite, Meteorites, Comets and the Moon are shown as half blue, half purple, half grey, half green and red symbols, respectively. The terrestrial and extraterrestrial data are taken from Ref. (*34-41*) and therein. Error bars shown in the figure are 1 σ, the error bar of Comet Coma is extraordinarily large (±100%, with n= 7401), other error bars are either not available or smaller than the symbols.

**Figure 4. Constraints on water fugacity $f(H_2O)$ by the crystallization of ULM-1.** The hexahydrate is stable above its thermodynamic equilibrium line (red), which constitutes the lower limit of $f(H_2O)$ for lunar volcanic gases. The vapor-liquid phase boundary of water forms the upper limits (blue). Between the red and gray lines, a dihydrate phase of NLM-1 become thermodynamically more stable. Reference to the formation temperature of the earth mineral (>100 °C), the lower limit of $f(H_2O)$ reached 1/3 of the record value in the driest volcano on earth - Lengai volcano (diamond). When considering $H_2$ degassing, the $f(H_2O + H_2)$ values become comparable to that of Lengai volcano (*47*) based on the $H_2/H_2O$ ratio estimated by Newcombe et al. (*19*, red dashed line) and our thermodynamic calculations (purple dashed line). $P^\Theta$ is the standard pressure (1 atm).